\documentstyle[12pt]{article}

\newcommand {\cD}{{\cal D}}

\def\a{\alpha}

\def\b{\beta}

\def\d{\delta}
\def\e{\epsilon}
\def\f{\phi}

\def\l{\lambda}
\def\m{\mu}
\def\n{\nu}
\def\o{\omega}
\def\p{\pi}
\def\q{\theta}
\def\r{\rho}
\def\s{\sigma}
\def\t{\tau}

\def\L{\Lambda}
\def\O{\Omega}

\newcommand{\vf}{\varphi}
\newcommand{\be}{\begin{equation}}
\newcommand{\ee}{\end{equation}}
\newcommand{\bea}{\begin{eqnarray}}
\newcommand{\eea}{\end{eqnarray}}

\begin{document}

\begin{titlepage}

\begin{flushright}
September 2010\\
\end{flushright}
\vspace{5mm}

\begin{center}
{\Large \bf  Nonlinear realizations of symmetries and unphysical Goldstone bosons
 }
\end{center}

\begin{center}

{\large  
I. N. McArthur\footnote{{mcarthur@physics.uwa.edu.au}}
} \\
\vspace{5mm}

\footnotesize{
{\it School of Physics, M013, \\
The University of Western Australia\\
35 Stirling Highway, Crawley W.A. 6009, Australia}}  
~\\

\vspace{2mm}

\end{center}
\vspace{5mm}

\begin{abstract}
\baselineskip=14pt
The embedding of a $p$-brane into higher dimensional spacetime breaks not only translational symmetries transverse to the worldvolume, but also Lorentz symmetries. There exist formulations for $p$-brane actions which associate Goldstone bosons with the generators of the broken Lorentz symmetries. These Goldstone bosons are unphysical, in that they  can be eliminated in favour of other Goldstone bosons either via their equations of motion or via the imposition of an inverse Higgs constraint. In this paper,  we examine the inter-relationship between the coset parameterization necessary to implement the inverse Higgs constraint, the equivalence of the inverse Higgs constraint to equations of motion, and the ability to find versions of the action with no explicit dependence on the unphysical Goldstone bosons. This is evidence that the unphysical Goldstone bosons are gauge degrees of freedom associated with an enlarged isotropy group. In addition to $p$-brane actions, a number of other cases, including conformally invariant dilaton actions, are shown to exhibit the same structure.

\end{abstract}
\vspace{1cm}

\vfill
\end{titlepage}

\section{Introduction}
Introduced by Ivanov and Ogievetsky \cite{Ivanov:1975zq}, the term ``inverse Higgs effect''  refers to the process  whereby a subset of the Goldstone bosons or gauge fields in a theory with nonlinearly realized symmetries can sometimes be eliminated algebraically in favour of the remainder by setting to zero a subset of the Cartan forms.

Here, we focus on the inverse Higgs effect as applied to Goldstone bosons associated with nonlinearly realized symmetries, and consider a large class of examples in which  certain of the Goldstone bosons are unphysical, in the sense that they can be eliminated algebraically in favour of other Goldstone bosons. We argue that this is because they are  gauge degrees of freedom. More specifically, the unphysical Goldstone bosons are degrees of freedom associated with an enlarged  isotropy group for the coset space related to the nonlinearly realized symmetries.

The  formulation of quantum field theories with nonlinearly realized symmetries was put on a systematic mathematical footing by Callan, Coleman, Wess and Zumino \cite{Coleman:1969sm, Callan:1969sn, Isham:1969ci}, and was extended to include spacetime symmetries by Volkov \cite{Volkov} (see also Ogievetsky \cite{Og}). The general structure is that a group $G$ of symmetries (that may include spacetime symmetries) is broken to a subgroup $H$ (which includes the Lorentz group in the case of theories with nonlinearly realized spacetime symmetries). Denoting the generators of $H$ by $\{ T_a \}$ and the remaining generators of $G$ by $\{ T_I \},$ we assume that the $ T_I$ provide a (possibly reducible) representation for the subgroup $H$ under the adjoint action,
$$
 [ T_a, T_b] = i \, f_{ab}{}^c\, T_c \, , \quad [ T_a, T_I ] = i \, f_{aI}{}^J \, T_J. 
 $$
No further assumptions about $G$ are made at this stage. 
 
We review the basic elements of the theory of nonlinear realizations to establish notation. The homogeneous space $G/H$ is  the set of left cosets in G under the equivalence relation $g \sim g h$ for $h \in H, $ and is usually parameterized locally by a ``slice'' through $G$ of the form
 $$ g (\xi) = e^{i \xi^I T_I}.$$ 
The left action of $G$ on G/H is realized via
 $$ g \, g(\xi) = g (\xi') \, h(g, \xi),$$
 where the compensating transformation by $h(g, \xi) \in H$ is in general necessary as $ g \, g(\xi)$ does not necessarily lie on the chosen slice through $G$. The subgroup $H$ is realized linearly on the $\xi^I.$ Of crucial importance are the Cartan forms $E^I, \Phi^a$ defined via the Cartan differential
 $$ g( \xi)^{-1} d g (\xi) = i (E^I \, T_I + \Phi^a T_a),$$
 which is invariant under the global left action of $G$ on $G/H.$ In field theories in which a global symmetry group $G$ is broken to a subgroup $H,$ the $\xi^I$ are Goldstone fields. Under the {\it local} right action $ g(\xi) \rightarrow g (\xi) \, h$ by $H$ (i.e.  $h \in H$ spacetime dependent), the components $E^I$ of the Cartan form transform covariantly and provide a vielbein on $G/H,$ whereas the components $\Phi^a$ provide a $H$-connection on $G/H.$
 
 As originally formulated \cite{Ivanov:1975zq}, the inverse Higgs effect applies in cases where there are both broken and unbroken spacetime symmetries \cite{Volkov, Og}. The symmetry groups $G$ of interest have generators $P_{\mu}$ of unbroken spacetime translations, generators $T_{i'}$ of broken symmetries (including broken spatial translations) associated with physical Goldstone bosons, further ``broken'' generators $T_{\alpha}$ associated with nonphysical Goldstone bosons (in the sense that they can be eliminated in favour of physical Goldstone bosons via the inverse Higgs effect), and generators $T_a$ of the unbroken symmetry group $H.$ The coset $G/H$ is parameterized in the form
 \be
 g = e^{i( x^{\mu}P_{\mu} + \xi^{i'} T_{i'})} \, e^{i \xi^{\a} T_{\a}},
 \label{param0}
 \ee
 where the Goldstone fields $\xi^{i'}$ and $\xi^{\a}$ are functions of the spacetime coordinates $x^{\mu}.$
 A necessary condition for implementation of the inverse Higgs effect  is that there are nonvanishing structure constants $f_{\mu \alpha}{}^{i'},$ as in this case, the Cartan forms $E^{i'}$ have the form
 $$ E^{i'} = d \xi^{i'} + \xi^{\a} d x^{\m} f_{\a \m}{}^{i'} + \cdots \, 
 =  \, d x^{\m} \, (\partial_{\m} \xi^{i'} +  \xi^{\a} f_{\a \m}{}^{i'} + \cdots), $$
 which allows the $\xi^{\a}$ to be eliminated algebraically in terms of derivatives of the $\xi^{i'}$ via the imposition of the inverse Higgs constraint $E^{i'} = 0.$ 
 

Inverse Higgs constraints have been used in the construction of  four general classes of action:  (i) Green-Schwarz actions in for $p$-branes embedded into $D$-dimensional Minkowski spacetime \cite{Hughes:1986dn, Hughes:1986fa, Townsend:1987yy, Gauntlett:1989qe, West:2000hr, Ivanov:1999gy, Low:2001bw} and in flat superspace  \cite{West:2000hr, Gomis:2006xw}; (ii) conformally invariant dilaton actions \cite{Salam:1970qk,  Isham:1970gz, Isham:1971dv, Borisov:1974bn, Ivanov:1988vw}, and the closely related  actions for $p$-branes embedded in $AdS_{p+2}$ space \cite{Kuzenko:2001ag, Delduc:2001tb, Bellucci:2002ji,  Ivanov:2002tb,  Ivanov:2003dt}; (iii) $AdS_d$-branes embedded in $AdS_{d+1}$ \cite{Clark:2005ht}; and  (iv)  actions with partially broken global supersymmetry (PBGS) \cite{Hughes:1986dn, Gauntlett:1989qe, Bagger:1994vj, Bagger:1996wp, Bellucci:2000bd, Bellucci:2000kc}.

Except for the case of PBGS, all of these examples have additional common structure, and it is the purpose of this paper to elucidate this structure. The first additional element is that the implementation of the inverse Higgs constraint is critically dependent on the explicit parameterization (\ref{param0}) of the coset $G/H.$ The second is that elimination of  the unphysical Goldstone bosons via their equations of motion yield actions which are the same as  those obtained via the implementation of the inverse Higgs constraint (as already observed by many authors). The final feature is that the actions obtained by elimination of the unphysical Goldstone bosons can be expressed via the Cartan forms in a manner whereby dependence on the unphysical Goldstone bosons drops out explicitly without any need to eliminate them via an inverse Higgs constraint or equations of motion. The aim of this paper is to provide an interpretation of all three of these characteristics in terms of the fact that the ``broken'' generators associated with the unphysical Goldstone bosons can be appended to the generators of the original isotropy group to form an enlarged isotropy group, thus showing that the corresponding Goldstone bosons are related to gauge degrees of freedom, which is why they are unphysical.
 The analysis in this paper does not apply to the case of actions with PBGS - these correspond to brane actions with worldvolume spinor degrees of freedom in addition to bosonic worldvolume degrees of freedom. The reason  PBGS actions cannot be treated is that there is no known systematic procedure to construct them from Cartan forms, as discussed in  \cite{Ivanov:1999fwa, Bellucci:2000bd, Bellucci:2000kc}.

There has been some prior work in these directions.  The fact that  imposing an inverse Higgs constraint yields the same action as elimination of a Goldstone boson by its equations of motion has been noted by many authors, including many of those referenced in cases (i) to (iv) above. In relation to enlargement of the isotropy group, Ivanov \cite{Ivanov:2003dt}  examined a mapping between dilaton actions in $p+1$ dimensional Minkowski space and $p$-brane actions in AdS space \cite{Bellucci:2002ji} via the use of surfaces in an extended conformal space. He points out that a way to restrict functions on this extended conformal space to appropriate subspaces is to require invariance under right shifts of the coset elements by an enlarged  isotropy group including the generators associated with unphysical Goldstone bosons (see Section 4 of \cite{Ivanov:2003dt}). Bellucci and Krivonos \cite{Bellucci:2006ts}, in the context of applying the techniques of nonlinear realizations to construct ${\cal N} = 4 $ superfields in one spacetime dimension, point out that setting some of the bosonic Goldstone fields to zero can enlarge the isotropy group by the corresponding generators (see section 3.4 of \cite{Bellucci:2006ts}). In the case of actions for $AdS_d$ branes propagating in $AdS_{d+1}$, Clark et al \cite{Clark:2005ht} have noted the existence of a Nambu-like version of the action in which the full symmetry of the action is manifest without the need to invoke the inverse Higgs effect. 

The outline of the paper is as follows. In section 2, simple examples in which unphysical Goldstone bosons are a feature of the construction of actions are examined. In all of these examples, elimination of the unphysical Goldstone bosons via their equations of motion is equivalent to the imposition of an inverse Higgs constraint. It is also the case that in all of these examples, there is a version of the action in which there is no explicit dependence on the unphysical Goldstone bosons, consistent with their interpretation as gauge degrees of freedom.  Section 3 contains a general discussion of the relationship between coset parameterization and the ability to consistently impose an inverse Higgs constraint. The issue of equivalence between inverse Higgs constraints and equations of motion for unphysical Goldstone bosons is addressed in section 4. Section 5 contains an analysis of dilaton actions for nonlinearly realized conformal symmetries in arbitrary spacetime dimensions - this requires special treatment as it does not fit with the general structure common to the other examples dealt with in this paper.

\section{Prototypical examples}
In this section, a number of simple examples of systems with nonlinearly realized spacetime symmetries are provided in which actions do not contain derivatives of certain Goldstone bosons. These Goldstone bosons are therefore unphysical as they can be eliminated algebraically via their equations of motion. In each case, these equations of motion are equivalent to  the imposition of an inverse Higgs constraint. There is also a version of the action in which there is no explicit dependence on the unphysical Goldstone without the need to impose equations of motion or inverse Higgs constraints, showing that the unphysical Goldstone bosons are gauge degrees of freedom.

\subsection{Relativistic particle in two-dimensions}
The first example is that of a a massive relativistic particle  moving in a two-dimensional spacetime with coordinates $(t,x),$ following the detailed treatment in  \cite{Gauntlett:1989qe} (see also \cite{Ivanov:1999gy}). The two-dimensional Poincar\'e algebra has generators $P_0,$ $P_1,$ and $M$ (respectively generators of temporal translations, spatial translations and boosts) satisfying the algebra
$$ [M, P_0] =  -i \, P_1, \quad [M, P_1] = - i \, P_0, \quad [P_0, P_1] = 0.$$
Embedding the worldline of the particle, parameterized by proper time $\tau,$ into the two-dimensional spacetime results in Goldstone bosons associated with  broken spacetime translations transverse to the worldline and  broken boost symmetry\footnote{The existence of a Goldstone field associated with the generator $M$ of boosts is based on the approach in which a $p$-brane embedded in $D$-dimensional flat spacetime is considered to break the group $ISO(D-1, 1)$ to  $SO(p,1) \times  SO(D-p-1),$ and so there are Goldstone bosons associated with broken Lorentz generators as well as the broken spacetime translations in directions transverse to the worldvolume \cite{Hughes:1986fa, West:2000hr, Ivanov:1999gy}. The Goldstone bosons associated with unbroken space-time translations can be eliminated by a choice of gauge associated with worldvolume diffeomorphism invariance.}.
Using standard techniques for theories with nonlinearly realized spacetime symmetries \cite{Volkov, Og}, the group is parameterized in the form\footnote{The isotropy group $H$ is trivial in this case. }
\begin{equation} 
g = e^{i t(\tau) P_0 +i  x (\tau) P_1} \, e^{i  \omega(\tau) M},
\label{param}
\end{equation}
where $\tau$ is the proper time parameterizing the worldline of the particle\footnote{ In  \cite{Gauntlett:1989qe}, static gauge is  adopted, so that $t = \tau.$}.
One finds
\begin{eqnarray}
& \, & E^0 P_0 + E^1 P_1 + \Phi M \nonumber \\ 
&=& -i \, g^{-1}d g \\
&=& (dt \cosh \omega - dx \sinh \omega) P_0 + (dx \cosh \omega - dt \sinh \omega ) P_1 + d \omega \, M, \nonumber
\label{vielbein}
\end{eqnarray}
where $\tau$ dependence of $t, \, x$ and $\omega$ has been suppressed.
The  constraint $E^1 = 0$ allows the Goldstone field $\omega$ to be eliminated algebraically,
\begin{equation}
 \omega =  {\rm arctanh} \,  \frac{dx}{d \tau}  \left( \frac{dt}{d \t} \right)^{-1}.
\label{phi}
\end{equation}
A natural reparameterization invariant action results from the integral of the $1+0$ dimensional einbein $E^0,$ 
\begin{equation}
 S_0 = - \, m \int E^0 = - \, m \int d \tau  \,( \frac{dt}{d \tau} \cosh \omega - \frac{d x}{d \tau}  \sinh \omega )
\label{S0}
\end{equation}
and eliminating $\omega$ via (\ref{phi}) yields the usual massive relativistic point particle action 
\begin{equation}
S_1 = - \, m \int d \tau \,  \sqrt{ \left( \frac{d t}{ d \tau} \right)^2  - \left(\frac{dx}{d \tau} \right)^2}.
\label{S1}
\end{equation}
Alternatively, starting with the action in (\ref{S0}), the equation of motion for $\omega$ turns out to be equivalent to the constraint $E^1 = 0,$ and so eliminating $\omega$ from the action (\ref{S0}) via its equation of motion also yields (\ref{S1}).

In relation to this construction, the first observation is that the elimination of the Goldstone field $\omega$ algebraically via the inverse Higgs constraint $E^1 = 0$ is critically dependent on the parametrization (\ref{param}).  To see this, consider the alternative parametrization
\begin{equation}
 g = e^{i (t P_0 + x P_1 + \omega M)}. 
 \label{altparam}
 \end{equation}
 In this case, one establishes  that 
 \begin{eqnarray}
 E^0 &=& dt + \frac{ (\cosh \omega - 1)}{\omega^2} \, (x \, d \omega - \omega \, d x) + \frac{( \sinh \omega - \omega)}{ \omega^2} \, ( t\,  d \omega - \omega \, dt ), \nonumber \\
 E^1 &=& dx + \frac{ (\cosh \omega - 1)}{\omega^2} \, (t \, d \omega - \omega \, d t) + \frac{( \sinh \omega - \omega)}{ \omega^2} \, ( x \, d \omega - \omega \, d x), \\
 \Phi &=& d \omega. \nonumber
 \end{eqnarray}
 It is now much more complicated to eliminate $\omega$ algebraically via an inverse Higgs constraint due to the $d \omega$ dependence of all the Cartan forms - complicated linear combinations of the Cartan forms must be taken to eliminate $d \omega $. In cases where there is a nontrivial isotropy group, the different Cartan forms will in general transform according to different representations of the isotropy group, and so it does not even make sense to take linear combinations of the Cartan forms if covariance of the theory under the action of the isotropy group is to be preserved.
 
 The second observation is that in the case of the parametrization (\ref{param}), the $\omega$ dependence appears in the  form of a right action, which is the same form as the local action of the isotropy group on representatives of a coset. This suggests that the reason $\omega$ can be eliminated as a physical degree of freedom is that it is a degree of freedom related to a generator of an enlarged isotropy group rather than a broken degree of freedom. If this is the case, there should be a version of the action in which the $\omega$ dependence of the action disappears without the need for imposition of a constraint or use of the equations of motion. And this is indeed so, since  the action
\begin{equation}
S_2 = - \, m \int \sqrt{ E^0 \otimes E^0 - E^1 \otimes E^1},
\label{S2}
\end{equation}
with $E^0$ and $E^1$ given by (\ref{vielbein}), is  independent of $\omega$ (using $\cosh^2 \omega - \sinh^2 \omega =1$) and equal to the action $S_1$ in (\ref{S1}). The action $S_2$ is none other than the Nambu action for a $p$-brane in the case $p=0,$ using the pullback  of the flat metric on two-dimensional spacetime to the worldline of the particle,  a formulation in which $SO(1,1)$ symmetry is linearly realized so that there is no need for a Goldstone boson associated with the breaking of boost symmetry. Invariance of the Nambu action under worldline diffeomorphisms ensures that there is only one physical Goldstone degree of freedom amongst $t(\tau)$ and $x(\tau)$, in that  $t(\tau)$ can be eliminated by a gauge choice  \cite{Townsend:1987yy}. Explicitly imposing the constraint $E^1 = 0$ also reduces the action $S_2$ to $S_0$ in (\ref{S0}).

The third observation we make relates to the fact that elimination of $\omega$ from $S_0$ in (\ref{S0}) via the inverse Higgs constraint $E^1 = 0$ yields an action equivalent to that obtained by elimination of $\omega$  from $S_0$ via its equation of motion. Using 
\begin{equation}
\delta (g^{-1} d g) = d v + [ \, g^{-1} d g, v \, ]
\label{variation}
\end{equation} 
with $v = g^{-1} \delta g,$ it follows from the parametrization (\ref{param}) that under a variation of $\omega,$
$$ v = i \, \delta \omega \, M.$$ Comparing (\ref{vielbein}) with (\ref{variation}), we find that under a variation of $\omega,$
\begin{equation} \delta E^0 = - \, \delta \omega \, E^1.
\label{deltaE0}
\end{equation}
Since $S_0$ is constructed from $E^0,$ $ \frac{\delta{S_0}}{\delta \omega} = 0$ implies via (\ref{deltaE0}) that $E^1 = 0,$ which is the inverse Higgs constraint. This is therefore the reason for the equivalence of the imposition of the inverse Higgs constraint and the elimination of $\omega$ via its equations of motion.  Again, this result is explicitly dependent on the parametrization (\ref{param}), and if $\omega$ is interpreted as parametrizing a degree of freedom in an enlarged isotropy (or gauge) group, it is a reflection of the fact that the action will contain only physical degrees of freedom and be independent of the gauge degrees of freedom associated with the isotropy group.

The more general case of the Nambu part of the action for $p$-branes embedded into $D$-dimensional Minkowski spacetime is a relatively straightforward extension of this $p=0, D=2$ example, and  the three features identified above are replicated  for general $p$ and $D.$ The coset parameterization is of the form
$$
 g = e^{i( x^{\mu}P_{\mu} + \xi^{i'} T_{i'})} \, e^{i \xi^{\a} T_{\a}},
$$
 where the $P_u$ generate worldvolume translations and (in physical gauge) the $x^{\m}$ are worldvolume coordinates, the $\xi^{i'}$ are Goldstone bosons associated with the $D-p-1$ broken translational symmetries transverse to the embedded $p$-brane, $H = SO(p,1) \times SO(D-p-1),$ and the $T_{\a}$ are the generators of the spacetime Lorentz group $SO(D-1,1)$ not contained in $H.$  The Goldstone bosons $\xi^{\a}$ associated with the broken Lorentz generators $T_{\a}$ are unphysical and can either be eliminated via their equations of motion or via the imposition of an inverse Higgs constraint, as in the example above. In this approach, the action is a worldvolume integral of the determinant of the vielbein obtained from the Cartan form, and elimination of the unphysical Goldstone bosons is required  in order to obtain the manifestly $SO(D-1,1)$ invariant Nambu action. The Goldstone bosons associated with the generators of space-time translations in the $p$-brane world volume can be eliminated by going to physical gauge (they are gauge degrees of freedom associated with worldvolume diffeomorphism invariance). 
 A physical discussion of reason why the Goldstone bosons $\xi^{\a}$ are unphysical in this case is contained in \cite{Low:2001bw}.
 
 As already noted in the introduction, these three features are characteristics of two other general classes of theory with broken spacetime symmetries for which actions are constructed via imposition of inverse Higgs constraints: (i) $p$-branes embedded in Minkowski space, flat superspace, and AdS spaces; and (ii) conformally invariant dilaton actions in arbitrary spacetime dimensions. Below we present additional simple well-known $1+0$ dimensional examples of each of these cases explicitly demonstrating the equivalence of the inverse Higgs effect to the equations of motion for an unphysical Goldstone boson, and also the existence of a formulation for the action in which dependence on the unphysical Goldstone boson drops out without any need to invoke an equation of motion or the inverse Higgs effect.

\subsection{Superparticle in two-dimensional superspace}
Here we consider a superparticle moving in two-dimensional flat superspace. The corresponding super-Poincar\'e algebra is 
\begin{eqnarray}
\, [M, P_0] &=& - i P_1, \quad [M, P_1] = - i P_0 \nonumber \\
\, [ M, Q_1] &=& - i Q_2, \quad [M, Q_2] = - i Q_1 \nonumber \\
\, \{ Q_1, Q_1 \} &=& \{ Q_2, Q_2 \} = - P_0  \\
\, \{ Q_1, Q_2 \} &=& - P_1. \nonumber
\end{eqnarray}
The embedding of the worldline of the superparticle, parameterized by proper time $\t,$ into the spacetime is parameterized in the form
$$g(\tau) = e^{i x^{\m}(\t)P_{\m} + i \q^{\a}(\t) Q_{\a}} \, e^{i \o(\t) M}.$$
The corresponding Cartan form is
$$g(\t)^{-1} d g(\t) = i \left(E^{\m}(\t) P_{\m} + {\cal E}^{\a}(\t) Q_{\a} + \O(\t)M \right), $$
where
\begin{eqnarray}
E^0 &=&  \p^0 \, \cosh \o -\p^1 \, \sinh \o, \quad E^1 =  \p^1 \, \cosh \o -  \p^0 \, \sinh \o\nonumber \\
{\cal E}^1 &=& d \q^1 \, \cosh \o -  d \q^2 \, \sinh \o, \quad {\cal E}^2  = d \q^2 \, \cosh \o -  d\q^1  \, \sinh \o \\
\O &=& d \o,
\end{eqnarray}
with $$ \p^{\m} = d x^{\m} - \frac{i}{2} \q \s^{\m} d \q$$
and with $\s^0 = {\bf 1}_2$ and $\s^1$ the usual Pauli spin matrix.
A natural candidate for a worldline reparameterization invariant action again involves the einbein $E^0:$
$$S = \int  E^0 = \int d \t \, \left(\p^0(\t) \, \cosh \o(\t) -\p^1(\t) \, \sinh \o (\t) \right).$$
Elimination of $\o(\t)$ via its equations of motion is equivalent to the inverse Higgs constraint $E^1 = 0$ since again $\frac{\delta E^0}{\delta \o} = - E^1,$ and the resulting action is the Nambu action
 $$ S = \int d \t \, \sqrt{ ( \p_{\t}^0)^2 - (\p_{\t}^1)^2}. $$
If instead we start with the Nambu action, explicit dependence of this action on $\o$ cancels out without the need to impose equations of motion or an inverse Higgs constraint.
 
 \subsection{Dilatons in one dimension}
The next example involves a conformally invariant dilaton action associated with spontaneous breaking of the conformal group 
$SO(1,2)$ of $1+0$ dimensional spacetime \cite{Ivanov:1988vw, Ivanov:2002tb, Ivanov:2003dt}.
In the standard or ``conformal'' basis \cite{ Ivanov:2002tb}, the Lie algebra of $SO(1,2)$ is
 \begin{eqnarray}
\, [ D, P] &=& - i \, P, \quad [D, K] = i \, K \nonumber \\
\, [ P, K ] &=& 2 i \, D.
 \end{eqnarray}
The coset is parameterized as
 $$ g = e^{i t P} \, e^{i u(t) D} \,e^{i \l(t) K}$$
 where $t$ is the time coordinate and  $u$ is the dilaton associated with spontaneously broken conformal symmetry.
 The Cartan form is
  $$ g^{-1} d g = i (\o_P P + \o_D D + \o_K K)$$
 with
 \begin{eqnarray}
 \o_P &=& e^{- u} dt, \quad \o_D = du - 2 e^{-u} \l dt \nonumber \\
 \o_K &=& d \l - \l du  + e^{-u} \l^2 dt.
 \end{eqnarray}
 
 This case differs from the two so far considered in that an action constructed from the einbein $\o_P$ has no dynamics. Instead, a nontrivial conformally invariant action for the dilaton degree of freedom is
  $$ S = - \int \omega_K. $$
 Eliminating $\l$ by its equation of motion is equivalent to imposing the inverse Higgs constraint $\o_D = 0,$ as $\frac{\delta \o_K}{\delta \l }= - \o_D,$  and the action becomes
 $$ S = \frac14 \int dt \, e^u \, \dot{u}^2.$$
 This action can also be written in the form 
 $$S = -  \int (\o_K - \frac14 \frac{\o_D{}^2}{\o_P}), $$ 
 in which explicit dependence on the unphysical Goldstone boson $\l$ drops out without the need for equations of motion or an inverse Higgs constraint.
This formulation of the action, which will be extended to higher dimensional conformally invariant dilaton actions in section 5, is related to work by Volkov \cite{Volkov}. Division by  $\o_P,$ which only makes sense in this one-dimensional case, is not required in higher dimensions. Also note that the action is not of ``Nambu-type,'' as it does not obviously involve the pullback to a worldline of a metric in a higher dimensional space.
 
 As shown in  \cite{Ivanov:2002tb,  Bellucci:2002ji, Ivanov:2003dt}, conformally invariant dilaton actions in $D$ dimensional Minkowski space can be related to the static gauge action for a $(D-1)$-brane embedded in $AdS_{D+1}$ \cite{Maldacena:1997re}.  In terms of nonlinear realizations, this requires a transition from the standard basis for the conformal group $SO(D,2)$ to the  ``$AdS$'' basis. In the case of $D=1,$ the conformal group is $SO(1,2),$ and the generators in the ``$AdS$'' basis are related to the standard basis by 
 $$ \hat{K} = mK - \frac{1}{m} P, \quad \hat{D} = m D.$$
 The Lie algebra in the $AdS$ basis takes the form
 \begin{eqnarray}
 \, [ \hat{D}, P ] &=& - i \,m P, \quad  [ \hat{D}, \hat{K} ] = 2 i \,P + i \,m \hat{K} \nonumber \\
 \, [P, \hat{K} ] &=& 2\, i \hat{D} .
 \end{eqnarray}
The coset parameterization is
$$g = e^{i t P} \,e^{i \f(t) \hat{D}} \,e^{i \O(t) \hat{K}}, $$
where $t$ is the worldline coordinate in static gauge.
The corresponding Cartan form is
$$ g^{-1} d g = i \,(\hat{\o}_P P +\hat{\o}_D \hat{D} + \hat{\o}_K \hat{K}).$$
With $ \L = \tanh \O,$
\begin{eqnarray}
 \hat{\o}_P &=& \frac{1 + \L^2}{1 - \L^2} e^{- m \f} dt - \frac{2 \L}{1 - \L^2} d \f \nonumber \\
  \hat{\o}_K &=& \frac{d \L}{1 - \L^2} + \frac{m \L}{1 - \L^2} (\L e^{- m \f} dt - d \f)  \nonumber \\
 \hat{\o}_D &=& \frac{1 + \L^2}{1 - \L^2} d \f  - \frac{2 \L}{1 - \L^2} e^{- m \f} dt
\end{eqnarray}
In this case, the action formed by integrating the einbein $\hat{\o}_P$ is nontrivial,
 $$ S = - \int \hat{\o}_P = \int dt \, \left( \frac{2 \L}{1 - \L^2} \partial_t \phi - \frac{1 + \L^2}{1 - \L^2}  e^{-m \f} \right). $$
The Goldstone field $\L$ appears without derivatives and can be eliminated via its equation of motion, which is equivalent to the inverse Higgs constraint $ \hat{\o}_D = 0$ via
 $$ \frac{\delta \hat{\o}_P }{\delta \L}  = - \frac{2}{1 - \L^2} \, \hat{\o}_D.$$ This allows  $\L$ to be expressed  in the form
 $$ \L = \frac{e^{m \f} \partial_t \f}{1 \mp \sqrt{1 - e^{2 m \f} (\partial_y \f)^2}},$$
 yielding the action
 $$S = \pm \int dt \, e^{- m \f} \sqrt{1 - e^{2 m \f} ( \partial_y \f)^2}. $$
Once again, this action can be expressed in a form in which explicit dependence on $\L$ drops out without the need to invoke an inverse Higgs constraint or impose equations of motion, namely
 $$ S = \pm \int dt \, \sqrt{ \hat{\o}_P{}^2 - \hat{\o}_D{}^2}.$$
 This is a Nambu-like action, in that it is related to the $SO(1,2)$ invariant metric on $AdS_2$ \cite{Ivanov:2002tb, Clark:2005ht}
 $$ ds^2 = - \hat{\o}_P^2 + \hat{\o}_D^2.$$
 Again, this construction generalizes to higher $D,$ whilst still retaining the features highlighted in the $D=1$ case above.

 
\section{Inverse Higgs constraints and coset \\
parametrization}

In this section, we show in a general context that the choice of coset parameterization plays a role in the ability to implement a consistent inverse Higgs constraint,  which is the first observation related to the example considered in section 2.
 
Suppose that the group $G$ is broken to the subgroup $H,$ and that the broken generators $\{T_I \}$ fall into two subsets, $\{T_i\}$ and $\{T_{\alpha} \},$ each subset transforming as a representation of the subgroup $H$. 
If we wish to have the potential  to establish algebraic relations expressing the Goldstone fields $\xi^{\alpha}$  associated with the generators $T_{\a}$ in terms of derivatives of the $\xi^i$  via an inverse Higgs constraint, then the $ \xi^{\alpha}$ must appear without derivatives in some subset of  the Cartan forms  $ \{ E^i, E^{\alpha} \}.$ Since the forms $E^{\alpha}$ contain $i \, d \xi^{\alpha} \, T_{\alpha},$ any inverse Higgs constraint must involve setting the vielbeins $E^i$ (or some subset of them if the $E^i$ transform reducibly under the action of the  isotropy group $H$) to zero. We do not address at this point the issue of the conditions under which the constraints can be solved to express the $\xi^{\a}$ in terms of derivatives of the $\xi^i.$

If we adopt the parametrization 
\be g = e^{i\xi^i T_i } \, e^{i \xi^{\alpha} T_{\alpha}}
\label{param1}
\ee
for $G/H$, then 
\begin{eqnarray}
i (E^i \, T_i +  E^{\alpha} \, T_{\alpha} \, + \omega^a \, T_a) 
&=&
g^{-1} d g \nonumber \\
&=& e^{-i \xi^{\a} T_{\a} }\, \left( e^{-i \xi^i T_i} \, de^{i \xi^j T_j} \right) e^{i \xi^{\beta} T_{\beta}} \nonumber \\
&\,& + \, e^{-i \xi^{\a} T_{\a}} \,de^{i \xi^{\beta} T_{\beta}} .
\label{altvielbein}
\end{eqnarray}
It is clear that all dependence on derivatives of $\xi^{\alpha}$ arises from the second term on the right hand side of (\ref{altvielbein}). Using 
\begin{eqnarray}
&\,&   e^{-i \xi^{\a} T_{\a}} \,d \, e^{i \xi^{\beta} T_{\beta}}  \nonumber \\
&=&  e^{-i \xi^{\a} T_{\a}}  \int_0^1 ds \, e^{i (1-s) \xi^{\beta} T_{\beta}} \, i \, d \xi^{\gamma} \, T_{\gamma} \, e^{i s \xi^{\d} T_{\d}} \nonumber \\
&=& i \sum_{n=0}^{\infty} \, \frac{ (-i)^n}{(n+1)!} \, ad^{(n)} (\xi^{\b }T_{\b}) \, (d \xi^{\a}T_{\a}),
\label{dbeta}
\end{eqnarray}
it follows that there can be no $d\xi^{\alpha}$ dependence in $E^i$ at order $n=1$ if the commutator $[T_{\beta}, T_{\gamma}]$ has no $T_i$ dependence i.e. $f_{\beta \gamma}{}^i = 0.$ Indeed, it is clear from the right hand side of  (\ref{dbeta}) that the absence of  contributions proportional to $d\xi^{\alpha}$ in $E^i$ {\it to all orders in $n$} requires that the set of generators $\{ T_{\alpha}, T_a\}$ close upon themselves under commutation i.e. that  $\{ T_{\alpha}, T_a\}$ generate a subgroup of $G.$ Given that the $T_{\alpha}$ transform as a representation of the subgroup $H,$ this is equivalent to the requirement that $f_{\alpha \beta}{}^i = 0.$ 
 
 The requirement that $\{ T_{\alpha}, T_a\}$ generate a subgroup of $G$ allows for the possibility of enlargement of the isotropy group from $H$ to this subgroup, in which case the $\xi^{\alpha}$ are gauge degrees of freedom corresponding to the generators $T_{\alpha}$ of the enlarged isotropy group. 
As a result, there is a formulation for the  action in which dependence on the unphysical Goldstone fields $\xi^{\a}$ drops out without the need to impose any constraints or equations of motion.  This is indeed what occurs in all of the cases presented  in section 2 , in both the one-dimensional examples presented there and their higher dimensional generalizations, as will be discussed in detail later in this section. The parameterization (\ref{param1}) is then self-consistent, in that the factor $e^{i \xi^{\a} T_{\a}}$ takes the form of a right action of an element of the enlarged isotropy group on the coset representative $g(\xi) = e^{i \xi^{i} T_{i}},$ and the field $\xi^{\a}$ is a gauge degree of freedom. Any constraint we impose by setting some of the Cartan forms to zero can be considered to be a (field dependent) gauge choice in this circumstance. 

On the other hand, if we adopt the parametrization
$$ g = e^{i(\xi^i T_i + \xi^{\alpha} T_{\alpha})}$$
of the coset $G/H,$ then
\begin{eqnarray}
i ( E^i \, T_i &+&  E^{\alpha} \, T_{\alpha} \, +  \omega^a \, T_a)  \nonumber \\
&=&
g^{-1} d g \nonumber \\
 &=& g^{-1} \, \int_0^1 ds \, e^{i(1-s) (\xi^j T_j + \xi^{\beta}T_{\beta})} \, \left(i \, d\xi^i T_i + i \, d \xi^{\alpha} T_{\alpha} \right) \, e^{ i s (\xi^k T_k + \xi^{\gamma}T_{\gamma})} \nonumber  \\
&=& i \, \sum_{n=0}^{\infty} \, \frac{ (-i)^n}{(n+1)!} \, ad^{(n)} \left( \xi^j T_j + \xi^{\beta}T_{\beta} \right) \,  \left( d\xi^i T_i + d \xi^{\alpha} T_{\alpha} \right) \nonumber  \\
&=& i \,  \sum_{n=0}^{\infty} \, \frac{ (-i)^n}{(n+1)!} \, ad^{(n)} \left( \xi^j T_j + \xi^{\beta}T_{\beta} \right) \,  \left( d\xi^i T_i \right) \nonumber  \\
& & + \,  i \, \sum_{n=0}^{\infty} \, \frac{ (-i)^n}{(n+1)!} \, ad^{(n)} \left( \xi^j T_j + \xi^{\beta}T_{\beta} \right) \,  \left(  d \xi^{\alpha} T_{\alpha} \right).
\label{fullviielbein}
\end{eqnarray}
To order $n=1,$ 
$$E^i = d \xi^i + \frac12 \xi^j d \xi^k \, f_{jk}{}^i + \frac12 (\xi^j d \xi^{\beta} - \xi^{\beta} d \xi^j ) \, f_{j \beta}{}^i + \frac12 \xi^{\beta} d \xi^{\gamma} f_{\beta \gamma}{}^i + \cdots.$$
Already at this order, absence of derivatives of $\xi^{\alpha}$ in $E^i$ requires\begin{equation}
 f_{j \beta}{}^i = 0,  \quad  f_{\beta \gamma}{}^i=0.
\label{constraint1}
\end{equation}
Alternatively, if the  $E^i$ transform reducibly under $H,$ an inverse Higgs constraint can involve setting some subset $E^{i'}$ of the vielbeine to zero (as is the case in the example in the previous section), in which case we require the less restrictive constraints
\begin{equation}
 f_{j \beta}{}^{\, i'} = 0,  \quad  f_{\beta \gamma}{}^i=0.
\label{constraint2}
\end{equation}
Either way, it no longer suffices to simply impose the requirement at the  $\{ T_{\alpha}, T_a\}$ close as an algebra, showing that the choice of parameterization (\ref{param1}) for $G/H$ is important if we wish to have the potential to use an inverse Higgs constraint to eliminate the fields $\xi^{\a}.$ Note that in   \cite{Low:2001bw}, the authors state that the number of physical Goldstone modes should be independent of the parameterization of the coset. However, the above analysis shows that a specific choice of parameterization is required to invoke the inverse Higgs mechanism.

To conclude this section, we come back to  the consistency requirement that broken generators $\{ T_{\alpha} \}$ associated with the unphysical Goldstone bosons close with the original isotropy group generated by the $\{T_a\}$ to generate a subgroup of $G$ which can be interpreted as an enlarged isotropy group. We show that this requirement is fulfilled in the higher dimensional generalization of all of the examples discussed in section 2. In the case of $p$-branes embedded in $D$-dimensional Minkowski space or flat superspace, the original isotropy group is $H = SO(p, 1) \times SO(D-p-1),$ with the remaining generators of the $D$-dimensional Lorentz group being broken by the embedding and associated with the unphysical Goldstone bosons. It is clear that these broken generators  combine with those of $H$ to generate $SO(D-1,1),$ which is the enlarged isotropy group associated with the Nambu form of the action. 

For conformally invariant dilaton actions in $D$-dimensional Minkowski space, the original isotropy group is  the $D$-dimensional Lorentz subgroup $H = SO(D-1,1)$ of the $D$-dimensional conformal group $SO(D,2).$ The generators associated with the unphysical Goldstone bosons are the generators $ K_{\m}$ of the special conformal transformations. It is clear that the $K_{\m}$ can be appended to the generators $M_{\m \n}$ of the Lorentz group to form an enlarged isotropy group, since they commute with each other and transform according to the vector representation of the Lorentz group\footnote{See section 5 for the full algebra of the conformal group.},
$$ [ M_{\m \n}, K_{\r} ] = \frac12 \eta_{\m \r} K_{\n} - \frac12 \eta_{\n \r} K_{\m}.$$ 
For the closely related case of $p$-branes embedded in $AdS_D,$ the isotropy group is still the  Lorentz subgroup $SO(D-1,1)$ of $SO(D,2),$ but the generators associated with the unphysical Goldstone bosons are the ``$AdS$ basis'' generators\footnote{See section 5 for details.}
$$ \hat{K}_{\m} = m K_{\m}  - \frac{1}{2m} P_{\m}.$$
Again these close with the $M_{\m \n}$ generators of the Lorentz group to form an enlarged isotropy group due to the fact that the $\hat{K}_m$ transform in the vector representation of the Lorentz group and satisfy the commutation relations
$$ [ \hat{K}_{\m}, \hat{K}_{\n} ] = - 4 M_{\m \n}.$$

\section{Equivalence of the inverse Higgs constraint to equations of motion} 
In this section, we consider the issue of equivalence between elimination of a Goldstone boson using an inverse Higgs constraint and elimination of the same field via its equations of motion. We are unable to carry out a completely general analysis as there does not appear to be a general form for the action in all such theories, as will be discussed in the next section. 

However, for a wide class of examples, namely the Nambu part\footnote{The Wess-Zumino term in the action does not contain the unphysical Goldstone bosons.} of the action for $p$-branes embedded in Minkowski spaces, flat superspaces, and AdS spaces, it is possible to prove the equivalence in general, because the action in all of these cases has the same form. The action is the determinant of a vielbein defined on the worldvolume of the $p$-brane from the Cartan form derived via the method of nonlinear realizations as applied to spacetime symmetries \cite{Volkov, Og}. This first, second and fourth examples in section 2 are the simplest cases of these actions.

The rest of this section is concerned with the proof of the equivalence of the inverse Higgs constraint and the equations of motion for the unphysical Goldstone bosons for this class of actions. We treat the case of conformally invariant dilaton actions separately in the next section, as the action takes a different form.


In all of the cases discussed in section 2 in which the worldvolume Lagrangian is the determinant of a vielbein on the worldvolume pulled back from the Cartan form for the nonlinearly realized symmetry, the coset parameterization is of the form (\ref{param0}). 
 \be
 g = e^{i( x^{\mu}P_{\mu} + \xi^{i'} T_{i'})} \, e^{i \xi^{\a} T_{\a}}.
 \label{param0}
 \ee
Here, $x^{\m}$ are coordinates on the brane in static gauge, the 
 $ \xi^{i'} $ are physical Goldstone bosons associated with broken generators of translations transverse to the worldvolume or dilatations, and the  $\xi^{\a} $ are unphysical Goldstone bosons associated with broken Lorentz transformations or special conformal translations.  The Cartan form is 
 $$ g^{-1} d g = i (E^{\m} P_{\m} + \o^{i'} T_{i'} + \o^{\a} T_{\a}),$$
 and the action is constructed from the vielbein 
  $E^{\m} = d x^{\n} \, E_{\n}{}^{\m}$  as
\be
S = \int d^{\,p+1} x \, \det E_{\n}{}^{\m} \equiv \int d^{\,p+1} x \, E.
\ee
This is invariant under the action of the isotropy group $H,$ as the vielbein transforms linearly under the action of this group.


We need to examine the structure of the equations of motion that result by varying with respect to the unphysical Goldstone bosons   $\xi^{\alpha}. $ The variation of the action is
\be
\delta S = \int d^{\, p+1} x \, E \, (E^{-1})_{\m}{}^{\n} \delta E_{\n}{}^{\m}. 
\ee
The variation of the vielbein follows from the variation of the Cartan form, which is 
\be
 \delta (g^{-1} d g) = d v + [ g^{-1} d g , v ]
 \label{deltaC}
 \ee
with 
\begin{eqnarray}  v &=&  g^{-1} \delta g \nonumber \\ 
&=&   e^{- i \xi^{\b} T_{\b}} \delta  e^{i \xi^{\alpha} T_{\alpha}} \nonumber \\
&=& i \, \delta \xi^{\alpha} T_{\alpha} + \cdots. 
\end{eqnarray} 
The higher order terms involve commutators of the $T_{\alpha}$ with themselves, and so closure of  $\{ T_{\alpha}, T_a \}$ as a subalgebra implies that $v$  contains only terms proportional to $T_{\alpha}$ and $T_a.$ 
We can write
$$ v = i \, \delta \xi^{\a} A(\xi)_{\a}{}^B  \, T_B,$$
where the $T_A$ denotes the generators of the enlarged isotropy group,  $T_{\alpha}$ and $T_a.$ 
As a result, the $dv$ piece in (\ref{deltaC})  does not contribute to $ \delta E^{\m},$
which  is determined by the $T_{\m}$ component of 
$$ [ g^{-1} d g, v ] = i \, [ E^{\m} T_{\m} + E^{i'} T_{i'} + E^{\alpha} T_{\alpha} + \Phi^a T_a, v ]. $$
Again using the closure of  $\{ T_{\alpha}, T_a \}$ as a subalgebra, only $ [ E^{\m} T_{\m} + E^{i'} T_{i'}  , v ] $  can potentially contribute to  $ \delta E^{\m}: $ 
$$
 \delta E^{\m} = i \, E^{\n} \, \delta \xi^{\a} \, A(\xi)_{\a}{}^B  \, f_{\n B}{}^{\m} +  i \, E^{i'} \, \delta \xi^{\a} A(\xi)_{\a}{}^B  \, f_{i' B}{}^{\m}.
$$
Using $ E^{\m} = dx^{\n} E_{\n}{}^{\m},$ 
$$ 
(E^{-1})_{\m}{}^{\rho} \, \delta E_{\rho}{}^{\m} = i \, \delta \xi^{\a} \, A(\xi)_{\a}{}^{B}  \, f_{\m B}{}^{\m} + i \, (E^{-1})_{\m}{}^{\rho} \, E_{\rho}{}^{i'} \, \delta \xi^{\a} \, A(\xi)_{\a}{}^B \, f_{i' B }{}^{\m}.
$$
Therefore, providing $f_{\m B}{}^{\m}$ is zero, the equation of motion is equivalent to the imposition of the inverse Higgs constraint $E^{i'} = 0,$ as the right hand side is proportional to $E_{\rho}{}^{i'}. $ The condition that $f_{\m B}{}^{\m}$ vanishes is the requirement that the matrices representing the action of the $T_B$ on the $P_{\m}$ under the adjoint action are traceless, which is true for the case of $p$-branes, as the $T_A$ generate the spacetime Lorentz group.


Another case, exemplified in the next section, is that in which the covariant action includes the components $E^{\a}$ of the Cartan form,  thereby introducing covariant derivatives of the $\xi^{\a}.$


 \section{Conformally invariant dilaton actions}
 In this section, we consider construction of conformally invariant dilaton actions using the inverse Higgs effect, corresponding to a nonlinear realization of the conformal group. This case differs in several respects from construction of brane actions in terms of Cartan forms, as will be pointed out below. We adopt the conventions\footnote{Note that the normalization of the generators in this section differs from that in section 2 by a factor of $i$.} used in \cite{Bellucci:2002ji, Ivanov:2003dt}.
 
The conformal group $SO(D,2)$ of $D$-dimensional Minkowski space has the algebra
 \begin{eqnarray}
\,  [ M_{\mu \nu} , M^{\rho \sigma} ] &=& \frac12 \,  \delta_{ \mu}{}^{\rho} M_{\nu }{}^{\sigma } - \frac12 \,  \delta_{ \n}{}^{\rho} M_{\m }{}^{\sigma } - \frac12 \,  \delta_{ \mu}{}^{\s} M_{\nu }{}^{\r}  + \frac12 \,  \delta_{ \n}{}^{\s} M_{\m }{}^{\r } \nonumber \\
\,  [M_{\m \n}, P{\rho} ] &=& \frac12  \, \eta_{\mu  \r} P_{\n} - \frac12 \, \eta_{\n  \r} P_{\m} \nonumber \\
\,  [ M_{\m \n}, K_{\r} ] &=& \frac12 \,  \eta_{\mu \r} K_{\n}  - \frac12 \,  \eta_{\n \r} K_{\m} \nonumber \\
\,  [ P_{\mu}, K_{\nu}] &=& 2( - \eta_{\m \n}D + 2 M_{\m \n} ) \nonumber \\
\,  [D, P_{\m}] &=& P_{\m} \nonumber \\
\,  [D, K_{\m} ] &= & - K_{\m},
 \end{eqnarray}
 where $\eta_{\m \n} = {\rm diag} (+, - , \cdots , - ).$
 The standard nonlinear realization of the conformal group \cite{Salam:1970qk,  Isham:1970gz, Isham:1971dv, Borisov:1974bn} involves choosing the Lorentz group $SO(D-1, 1)$ of Minkowski space as the stability group, and parameterization of the coset in the form
 \be
 g = e^{x^{\m} P_{\m}} \, e^{\vf D} \, e^{\Omega^{\n} K_{\n}}.
 \label{paramconf}
 \ee
 The Cartan forms are computed to be \cite{Bellucci:2002ji, Ivanov:2003dt}
 \be
 g^{-1} d g =i (  \omega_P^{\m} P_{\m} + \omega_D D + \omega_K^{\m} K_{\m} + \o_M^{\m \n} M_{\m \n} )
 \ee
 with
 \bea
\,  \o_P^{\m} &=& e^{- \vf} \, dx^{\m} \nonumber \\
\, \o_D &=& d \vf - 2 e^{- \vf} \O_{\m} dx^{\m} \nonumber \\
\, \o_{K}^{\m} &=& d \O^{\m} - \O^{\m} d \vf + e^{- \vf}  \O_{\n}  (2 dx^{\n} \O^{\m} -\O^{\n} dx^{\m}) \nonumber \\
\, \o_M^{\m \n } &=& - 2 e^{- \vf} (\O^{\m} dx^{\n} - \O^{\n} dx^{\m} ).
\eea
 The inverse Higgs constraint $\o_D = 0$ allows the Goldstone field $\O_{\m}$ to be eliminated via
 $$ \O_{\m} = \frac12 e^{\vf} \, \partial_{\m} \vf.$$
 A covariant action is constructed in the form 
 \bea S &=&  \int d^{d}x \, e^{- d \vf} \, \cD_{\m} \O^{\m} \nonumber \\
 &=&   \int d^{d}x \, e^{- (d-1) \vf} \left( \partial_{\m} \O^{\m} - \O_{\m} \partial^{\m} \vf -(d-2) \e^{- \vf} \O_{\m} \O^{\m} \right)
 \label{Svf}
 \eea
 where $\o_K^{\m} = \o_P^{\n} \, \cD_{\n} \O^{\m}.$
 Eliminating $\O_{\m}$ via the inverse Higgs constraint yields
  $$ S = \frac14 \,  (d-2) \int d^{d}x \, e^{-(d-2) \vf}  \, \partial^{\m} \vf \partial_{\m} \vf. $$
  This same action can also be obtained by eliminating $\O_{\m}$ from (\ref{Svf}) via its equations of motion.

In order to investigate the equivalence of the inverse Higgs constraint and the equations of motion, it is convenient to note that the action (\ref{Svf}) can be written in the form 
\be
S_0 = \frac{1}{(d-1)!} \,\int \epsilon_{\m_0  \cdots \m_{d-1}} \, \o_K^{\m_0} \wedge  \o_P^{\m_1} \wedge \cdots \wedge \o_P^{\m_{d-1}}.
\ee
The appearance of $ \o_K^{\m}$ in the construction of this action places it in a different class to the $p$-brane actions, which involve only Cartan forms associated with generators of  spacetime translations (i.e. the analogue of  $\int \epsilon_{\m_0  \cdots \m_{d-1}} \, \o_P^{\m_0} \wedge   \cdots \wedge \o_P^{\m_{d-1}}$ ).
If we vary $\O^{\m},$
\bea
\delta S_0 &= &  \frac{1}{(d-1)!} \,\int \epsilon_{\m_0  \cdots \m_{d-1}} \, \delta \o_K^{\m_0} \wedge  \o_P^{\m_1} \wedge \cdots \wedge \o_P^{\m_{d-1}} \nonumber \\
& & + \int \epsilon_{\m_0  \cdots \m_{d-1}} \,  \o_K^{\m_0} \wedge  \delta \o_P^{\m_1} \wedge \cdots \wedge \o_P^{\m_{d-1}}. 
\eea
Using (\ref{deltaC}), one computes
\bea
\, \delta \o_P^{\m} &=& 0 \nonumber \\
\, \delta \omega_D  &=& - 2 \, \delta \O_{\m} \, \o_P^{\m} \nonumber \\
\, \delta \omega_K^{\m} &=& d (\delta \O^{\m}) - \delta \O^{\m} \, \o_D + \delta \O^{\n} \,  \o_{M \n}{}^{\m} \nonumber \\
&\equiv &  D \delta \O^{\m} - \delta \O^{\m} \, \o_D  \nonumber \\
\, \delta \o_M^{\m \n} &=& -  \,\delta \O^{\m} \,  \o_P^{\n} +  \delta \O^{\n} \, \o_P^{\m},
\eea
where $D$ is the covariant derivative defined by the connection form $\o_{M\nu}{}^{\mu}.$ 
This yields
\be
\delta S_0 =  \frac{1}{(d-1)!} \,\int \epsilon_{\m_0  \cdots \m_{d-1}} \, \left( D \,\delta \O^{\m_0} - \delta \O^{\m_0} \, \o_D  \right) \wedge  \o_P^{\m_1} \wedge \cdots \wedge \o_P^{\m_{d-1}}.
\ee
Integrating by parts and using $D \o_P^{\m} \equiv d \o_P^{\m}  - \o_P^{\n} \wedge \o_{M \n}{}^{\m} = -  \o_D \wedge \o_P^{\m},$
\be
 \delta S_0 = \frac{(d-2)}{(d-1)!} \, \int  \epsilon_{\m_0  \cdots \m_{d-1}} \, \delta \O^{\m_0} \, \o_D \wedge \o_P^{\m_1} \wedge  \cdots \wedge \o_P^{\m_{d-1}},
\label{deltaS}
\ee
thus establishing that the inverse Higgs constraint $\o_D = 0$ is equivalent to the equation of motion $\frac{\delta S_0}{\delta \O_{\m}} = 0.$

If it is possible  to find another combination of Cartan forms whose variation exactly cancels that in (\ref{deltaS}), then we can formulate an action that is manifestly independent of $\O_{\m}$ (this is a little reminiscent of anomaly cancellation by addition of a counterterm). This would be consistent with an enlargement of the isotropy group to include the generators $K_{\m},$ since the generators $\{K_{\m}, M_{\n \r} \}$ form a closed algebra. Using the fact that $\o_D$ defines a covariant derivative of the Goldstone field $\vf$ via $\o_D = \o_P^{\n} \, \cD_{\n} \vf,$ we can write
\bea
\delta S_0 & = & \frac{(d-2)}{d!} \, \int \delta \O^{\n} \, (\cD_{\n} \vf) \, \epsilon_{\m_0  \cdots \m_{d-1}} \,  \o_P^{\m_0} \wedge  \cdots \wedge \o_P^{\m_{d-1}} \nonumber \\
&=& - \frac14 \, \frac{(d-2)}{d!} \, \, \delta  \int \left( \cD^{\n} \vf \, \cD_{\n} \vf \right) \, \epsilon_{\m_0  \cdots \m_{d-1}} \,  \o_P^{\m_0} \wedge  \cdots \wedge \o_P^{\m_{d-1}},
\eea
where we have used $ \delta \o_D = - 2 \, \delta  \O_{\m} \, \o_P^{\m}$ and $\delta \o_P^{\m} = 0.$
Defining 
$$
S_1 = \frac14 \, \frac{(d-2)}{d!} \, \int \left( \cD^{\n} \vf \, \cD_{\n} \vf \right) \, \epsilon_{\m_0  \cdots \m_{d-1}} \,  \o_P^{\m_0} \wedge  \cdots \wedge \o_P^{\m_{d-1}},
$$
the action $S_0 + S_1$ has  no manifest $\O_{\m}$ dependence, as can easily  be checked by explicit calculation.
If the $\o_P^{\m}$ are taken to be vielbeins for a conformally flat metric $e^{- 2 \vf} \eta_{\m \n},$
$$ * \, \o_D = \frac{1}{(d-1)!} \,  \epsilon_{\m_0  \cdots \m_{d-1}} \, ( \cD^{\m_0} \vf ) \, \o_P^{\m_1} \wedge \cdots \wedge  \o_P^{\m_{d-1}},$$
and the action $S_1$ can be expressed in the compact form
$$
S_1 = \frac{(d-2)}{4} \, \int \o_D \wedge * \, \o_D.
$$
Note that the action $S_0 + S_1$ is not of ``Nambu'' form, in that it contains two separate pieces, which also places the dilaton actions in a different class to brane actions in terms of their construction via Cartan forms. Related actions have been considered by Volkov \cite{Volkov}.

\section{Conclusion}
In this paper, we have explored the reasons underlying the frequently observed phenomenon whereby elimination of nonphysical Goldstone bosons via their equations of motion is equivalent to an inverse Higgs constraint. The  inverse Higgs constraint is related to a particular choice for coset parameterization in which the unphysical Goldstone bosons appear in the form of a right action on the coset representative, suggestive of the fact that it may be associated with a gauge degree of freedom via an enlargement of the isotropy group. The existence of versions of the action in which symmetry under the enlarged isotropy group is manifest and therefore the nonphysical Goldstone bosons make no explicit appearance is consistent with this interpretation.

The analysis in this paper applies only in cases where actions are constructed from the Cartan form. There is no known systematic procedure for construction of actions with partially broken global supersymmetry from the Cartan form, although such actions are frequently constructed by invoking an inverse Higgs constraint to eliminate unphysical Goldstone bosons. It would be of interest to understand whether the unphysical Goldstone bosons in this case can also be related to an enlargement of the isotropy group. \\
\\

\noindent
{\bf Acknowledgements} \\
The author is grateful to  Sergei Kuzenko for helpful discussions and  critical feedback on the manuscript.

\end{document}